\documentclass[12pt,preprint]{aastex}
\usepackage{ulem}

\newcommand{\mvir}{M_{\rm vir}}
\newcommand{\rvir}{r_{\rm vir}}

\newcommand{\reffvirtot}{r_{\rm e, h}^{\rm tot}}
\newcommand{\sigmavirial}{\sigma_{\rm h}^{\rm tot}}
\newcommand{\mbarvir}{M_{\rm h}^{\rm bar}}
\newcommand{\mcbvir}{M_{\rm h}^{\rm cb}}
\newcommand{\mhbvir}{M_{\rm h}^{\rm hb}}
\newcommand{\cf}{c_{\rm F}}
\newcommand{\cmvir}{c_{\rm M}^{\rm vir}}
\newcommand{\stellarmass}{M_{\rm bo}^{\rm star}}

\newcommand{\restar}{r_{\rm e, bo}^{\rm star}}
\newcommand{\sigmastar}{\sigma_{\rm bo}^{\rm star}}
\newcommand{\cfstar}{c_{\rm F}^{\rm star}}

\slugcomment{\today}

\shorttitle{Massive galaxies at High-$z$}
\shortauthors{O\~norbe et al.}

\begin{document}


\title{Massive Galaxies at High-$z$: Assembly Patterns, Structure \& Dynamics in the Fast Phase of Galaxy Formation}


\author{J. O\~norbe\altaffilmark{1,2}, F.J. Mart\'inez-Serrano\altaffilmark{3}, R. Dom\'inguez-Tenreiro\altaffilmark{1}, A. Knebe\altaffilmark{1} and A. Serna\altaffilmark{3}}
\altaffiltext{1}{Grupo de Astrof\'isica, Departamento de F\'isica Te\'orica, Modulo C-15, Universidad Aut\'onoma de Madrid, Cantoblanco E-28049, Spain}
\altaffiltext{2}{Department of Physics and Astronomy, University of California at Irvine, Irvine, CA 92697, USA}
\altaffiltext{3}{Depto. de F\'isica y A.C., Universidad Miguel Hern\'andez, E-03202 Elche, Alicante, Spain}
\email{jonorbeb@uci.edu}


\begin{abstract}
 Relaxed, massive galactic objects  have been identified at redshifts $z=4, 5,$ and $6$
in hydrodynamical simulations run in a large cosmological volume.
 This allowed us to analyze the assembly patterns  of the high mass end of the
 galaxy distribution at these high $z$s, by focusing on their structural and dynamical
properties. Our simulations indicate that massive
objects at high redshift already follow
certain scaling relations.  These relations define virial planes at the halo scale,
whereas at the galactic  scale they
define {\it intrinsic dynamical planes} that are, however, tilted relative to
the virial plane. Therefore, we predict that massive galaxies must lie on fundamental
planes from their formation. 

We briefly discuss the physical origin of the tilt in terms the physical
processes underlying massive galaxy formation at high $z$, in the context of a
two-phase galaxy formation scenario. Specifically, we have found that it lies on
 the different behavior of the
gravitationally heated gas as compared with cold gas previously  involved in caustic
formation, and the mass dependence of the energy available to heat the gas.

\end{abstract}

\keywords{ galaxies: elliptical and lenticular, cD  --- galaxies: Evolution ---
galaxies: Formation --- galaxies: Fundamental Parameters --- hydrodynamics ---
methods: numerical}

\section{Introduction}
\label{s1:intro}

 One of the outstanding yet most important problems
  in astrophysics is how and when galaxies formed within the framework
  of the expanding Universe described by the concordance model of
  cosmology. 
Massive galaxies at high $z$ become more and more important to study.
In fact, the availability of multi-wavelength
  data from new generations of deep surveys, including wide field panoramic surveys, allowed for searches for
  such massive galaxy candidates up to $z \simeq 4 - 6.5$
\citep{Giavalisco:2004,Mobasher:2005,McLure:2006,
    Yan:2006,Rodighiero:2007,Bouwens:2007,
Wiklind:2008,Mancini:2009,Stark:2009,Mobasher:2010,Dahlen:2010,Capak:2011} or even $z \approx 10$ \citep{Bouwens:2011}.
  However, still very few is known about the physical processes
  underlying the (putative) presence of such massive systems at these
  high redshifts. In fact, the mere existence of them could seem
  paradoxical within a direct interpretation of the hierarchical
  structure formation scenario
  \citep[e.g.][]{Toomre:1977,White:1978}.  Further, such possible
  contradictions are not necessarily alleviated by the competing
  monolithic collapse scenario
  \citep{Eggen:1962,Larson:1974} and hence the question
  about the existence (and the properties) of massive galaxies at
  high-redshift remains open.

Recently, a scenario has emerged  to explain massive galaxy
  formation that shares characterisctics of both the aforementioned
  classical scenarios, but is nevertheless different. Indeed,
analytical models \citep{SalvadorSole:2005}, as well as N-body
simulations \citep{Wechsler:2002,Zhao:2003}, have shown that two
different phases can be distinguished along {\it halo} mass assembly:
i) first, a violent, fast phase, with high mass aggregation (i.e.,
merger) rates, ii) later on, a slow phase, where the mass aggregation
rates are much lower. Hydrodynamical simulations have confirmed this
scenario and its implications for properties of massive galactic
objects at low $z$, see \citet{DT:2006}, see also \citet{Oser:2010}
and \citet{Cook:2009}.
Concerning high $z$s, it has been shown that the fast phase has the
characteristics of a {\it multiclump collapse}, where mergers involve
very low relative angular momentum, and, in fact, they are induced by
the collapse of flow convergence regions displaying a web-like
morphology \citep{DT:2010}.

In this \textit{Letter} we investigate the high mass end of galactic
stellar objects at high-redshift ($z=4$, $5$ and $6$)  
  obtained by means of self-consistent cosmological simulations within
  a volume large enough to account for the proper treatment of the
  large-scale structure yet simultaneously capturing all the relevant
  small-scale (baryonic) physics. 
  Not only do we investigate their mere presence, we also
  study whether they  
  had enough time to dynamically relax at such high redshifts.  To
this end, we have focused on the intrinsic mass (as
opposed to luminosity) as well as structural and kinematical
properties of these objects at their halo (i.e. virial radius) and
stellar/galactic scale (to be defined below) as fingerprints of the
physical processes involved in their assembly.\footnote{
  We stick to 3D properties as well as mass (instead of luminosity)
  for two reasons: i) projection effects add noise in the statistical
  analysis \citep{Onorbe:2006} and ii) we are not aiming at providing
  observables but rather at understanding the physical processes
  involved in the formation of these objects.} 
%Please note that it is
%  yet not even possible to observationally establish the scaling
%  relations reported here and hence we do not aim at such a
%  comparison.} 
Specifically, we investigate the appearence of samples
of high-$z$ massive galactic objects with dynamical planes tilted
relative to the virial plane, and link their underlying
  formation physics to the Adhesion Model
\citep{Gurbatov:1989,Vergassola:1994}.

Our results here are an extension to higher $z$s of previous studies
on the fundamental plane (FP) at $z=0$
\citep{Onorbe:2005,Onorbe:2006}.   In these two
  papers the different possibilities causing the tilt of the FP
  relative to the virial plane are analyzed in detail. It is shown
  that that if both the virial mass to luminosity ratio, $\mvir/L$,
  and the mass structure coefficient $\cmvir$
  (see Eqs. 2 and 4 in \citet{Onorbe:2005}) are independent of mass,
  then no FP tilt would be measured.  Here, because mass is considered
  instead of luminosity and no projection effects are taken into
  account, we have instead analyzed the mass dependence of the ratio
  $\stellarmass/\mvir$ and the mass structure
  coefficient $\cmvir = (G \mvir)/([\sigmastar]^{2}
  \restar)$, where $\stellarmass$ and $\sigmastar$ are the
  stellar mass and the 3D stellar velocity dispersion at the galactic
  scale, respectively, and $\restar$ is the 3D stellar mass effective
  radius also defined at the galactic scale\footnote{Please refer to
    Table 1 of \citet{Onorbe:2006} where our nomenclature and
    definitions are more thoroughly introduced.}. We need to stress
  that a mass dependence of either of these quantities automatically
  implies a tilt of the dynamical plane relative to the virial plane.

\section{Structural and kinematical Properties of massive Objects at high-$z$}
\label{s2:method}

The simulations used here are part of the GALFOBS
project. They are $N$-body $+$ SPH
simulations that have been performed using an OpenMP parallel version
of the \texttt{DEVA} code \citep{Serna:2003} and the methods for star
formation and cooling described in \citet{MartinezSerrano:2008}. The
\texttt{DEVA} code pays particular attention to ensure that conservation laws
(e.g. momentum, energy, angular momentum, and entropy) hold as
accurately as possible\footnote{This in particular implies that a
  double loop in the neighbour searching algorithm must be used, which
  considerably increases the CPU time.}. Star formation is implemented
through a Kennicutt-Schmidt-like law with a density threshold
$\rho_{thres}$ and a star formation efficiency of $c_{∗}$. The values
of these parameters implicitly account for star formation regulation
by discrete energy injection processes.

The main simulation was carried out in a periodic cube of 80 Mpc side
length using $512^{3}$ baryonic and $512^{3}$ dark matter particles
with a gravitational softening of $\epsilon_{g}=2.3$ kpc and a minimum
hydrodynamical smoothing length half this value. The cosmology applied
was a $\Lambda$CDM model whose parameters as well as those of the
field of primordial density fluctuations (i.e., initial spectrum) have
been taken from CMB anisotropy data\footnote{
  http://lambda.gsfc.nasa.gov/product/map/current/params/lcdm\_sz\_lens\_run\_wmap
  5\_bao\_snall\_lyapost.cfm} \citep{Dunkley:2009}, with
$\Omega_{m}=0.295$, $\Omega_{b} =0.0476$, $\Omega_{\Lambda}=0.705$,
$h=0.694$, an initial power-law index $n=1$, and
$\sigma_{8}=0.852$. The mass resolution is $m_{bar}=2.42\times10^{7}
M_{\odot}$ and $m_{dm}=1.26\times10^{8} M_{\odot}$ and the star
formation parameters used were $\rho_{thres}=4.79\times10^{-25} g \dot
cm^{-3}$ and $c_{∗}=0.3$.

When analyzing galaxy formation in numerical simulations it is
desirable to verify that the objects in the simulation are consistent
with observations at low $z$'s.  Due to the extreme CPU consumption by
hydrodynamical forces, this is not yet possible for the main GALFOBS
simulation. As a way out, we ran three sub-volumes of the main cube
using a ``zoom approach''. In this approach the gravitational forces
have been calculated for the full box whereas the hydrodynamical
forces (which are exclusively local) were only computed in a sub-box
of side length 26 Mpc.  These three sub-volumes have been analysed at
redshift $z=0$ showing that we indeed obtain galaxy populations in
agreement with low-redshift observations,  as we had also
  previously shown in \citet{Saiz:2004} using the same approach yet
  smaller simulation boxes.

Halos in our simulations are identified by the OpenMP+MPI halo finder
\texttt{AHF}\footnote{\texttt{AHF} can be freely downloaded from
  \texttt{http://popia.ft.uam.es/AMIGA}} \citep{Knollmann:2009} as
well as \texttt{SKID} \citep{Weinberg:1997},
and their respective results  have been
cross-compared to check for completeness.
  The halo scale of these
objects is defined by the virial radius ($\rvir$) based upon the
\citet{Bryan:1998} fitting function to determine the overdensity
threshold.  The so-called galactic scale has been based upon material (stars)
inside a sphere of radius $r=0.15\times \rvir$, a scale separating the
baryon from the dark matter domination \citep{Bailin:2005}. 
This automated procedure has been tested by comparing with individually determined  limiting
stellar sizes of several hundred of objects based upon their 3D visualization
as well as their 3D stellar density profiles.
We further asked that our objects are not involved in violent events,
either at the halo or at the galactic scale.
To exclude this kind of objects at the halo scale, we have used the form factor, $\cf$, defined via the virial
relation $\cf = (G \mvir)/([\sigmavirial]^{2} \reffvirtot)$, where
$\sigmavirial$ is the velocity dispersion and $\reffvirtot$ the
half-mass radius at the halo scale, and we asked it to be 
within the expected interval (1.9, 2.5) for virialized
objects \citep{Binney:2008} and of the order of unity if we use
$\rvir$ instead of $\reffvirtot$.
The same procedure has been employed on the galactic object
scale using -- in analogy -- the parameter $\cfstar = (G
\stellarmass)/([\sigmastar]^{2} \restar)$.
 Again,
objects outside a certain range (based upon a manually gauged
subsample of 200 objects for each $z$) have been discarded. 
Putting a mass threshold of $\stellarmass > 10^{10} M_{\odot}$, our final samples
 consists of 137, 521  and 1315  galaxies at
$z=6$, $z=5$ and $z=4$, respectively, not involved in violent events at any scale.

Our first result is in fact the mere existence of these samples of
high-redshift relaxed massive galaxies.  To understand their
origin, we first quantify the correlation and inter-relation,
respectively, between their mass ($\mvir$ and $\stellarmass$), size
(half-mass radii $\reffvirtot$ and $\restar$) and velocity 
dispersion\footnote{We stress that all our objects here are velocity dispersion supported}
($\sigmavirial$ and $\sigmastar$) both at the halo and at the stellar
scale using the following variables: $E_{halo} \equiv \log_{10}
\mvir$, $r_{halo} \equiv \log_{10} \reffvirtot$, $v_{halo} \equiv
\log_{10} \sigmavirial$, and $E_{star} \equiv \log_{10} \stellarmass$,
$r_{star} \equiv \log_{10} \restar$, $v_{star} \equiv \log_{10}
\sigmastar$.  We list the average values $\textit{\~E}$,
$\textit{\~r}$, and $\textit{\~v}$ in Table~\ref{tab:PCA} where we can
observe a mild increase of $r_{halo}$ alongside a decrease of
$v_{halo}$ while $E_{halo}$ remains constant: as the Universe expands,
the objects become on average less and less compact due to the
decrease of the global density \citep{Padmanabhan:1993}. We basically
observe the same phenomenon on the stellar scale, however, accompanied
by a moderate increase in $E_{star}$.  In fact, the ratios $
\reffvirtot/ \restar$ and $\sigmavirial/\sigmastar$ show an scaling
behaviour as a function of either $\mvir$ or $\stellarmass$ (see Table
2 and below).

Going one step further,
we search for planes in the ($E,r,v$) space by performing a principal component
analysis (PCA) of all samples. It is made in 3D to circumvent projection effects
\citep{Onorbe:2006}. We have found that at all redshifts one of the eigenvalues
of the PCA is considerably smaller than the others, 
so that (massive) objects populate a flattened ellipsoid close to 
two-dimensional, 
 both at the halo scale 
and  at the stellar object scale:
\begin{equation}
E_{s} - \textit{\~E}_{s} = \alpha^{\rm 3D}_{s} (r_{s} - \textit{\~r}_{s}) +
\gamma^{\rm 3D}_{s} (v_{s} - \textit{\~v}_{s}),
\label{eq:planeEq}
\end{equation}
where $s$ refers to the scale of the object, i.e. halo or star.
Table~\ref{tab:PCA}  also lists the values of the parameters $\alpha^{\rm 3D}_{s}$
and $\gamma^{\rm 3D}_{s}$ of these planes as well as their bootstrapping errors.
We find that at the halo scale the planes are close to the virial plane (VP,
defined by $(\sigmavirial)^{2} = G \mvir/\reffvirtot$ or
$ \alpha^{\rm 3D}_{halo} = 1, \gamma^{\rm 3D}_{halo}= 2$), 
as expected for well-defined haloes. At the stellar scale we also find
planes
to which we refer as the \textit{intrinsic dynamical planes (IDPs)} and whose
observed manifestation is the Fundamental Plane.

\begin{figure}
\plotone{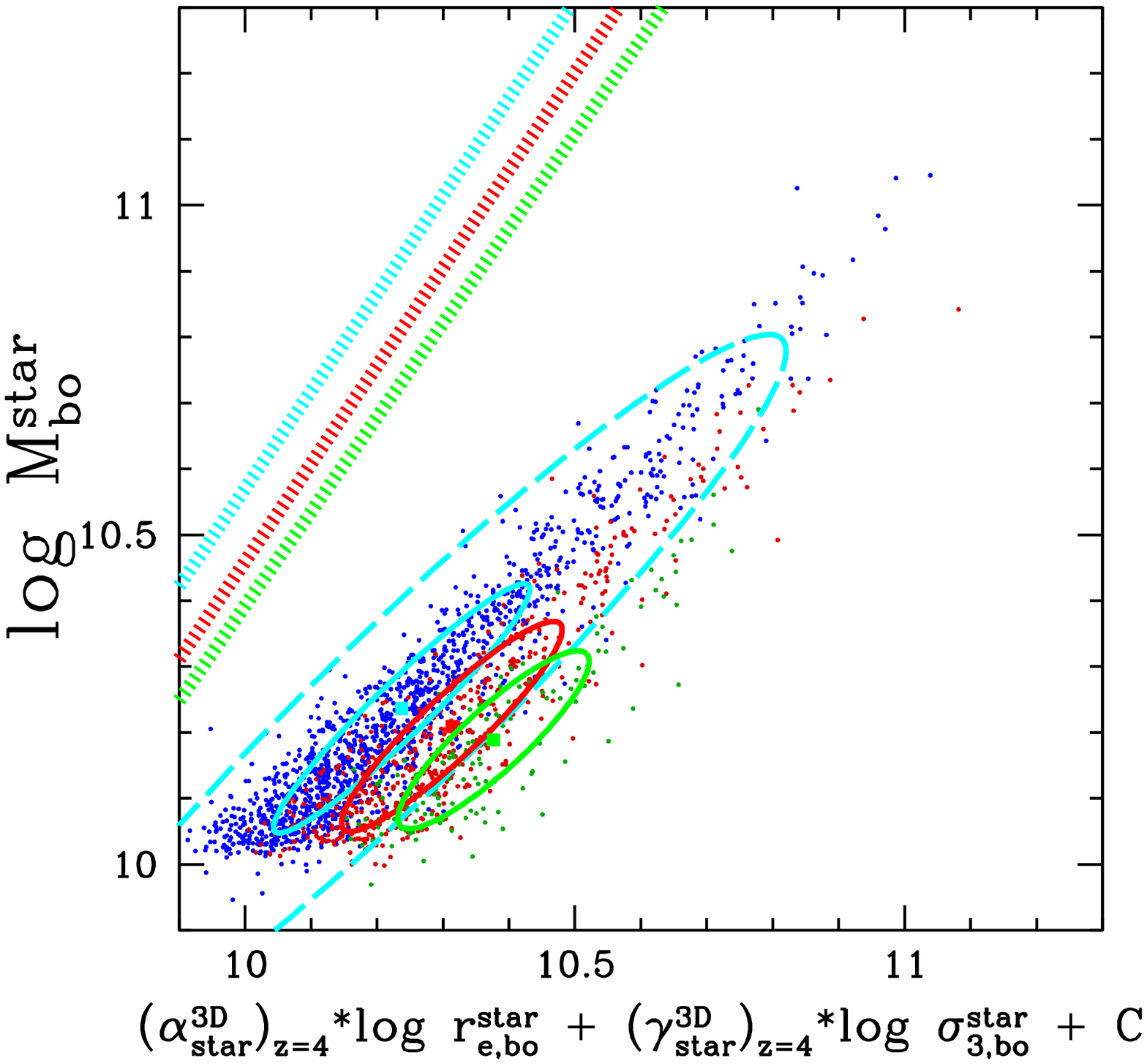}
\caption{The $z=4$ (blue), $z=5$ (red) and $z=6$ (green) IDPs
seen in a projection where the $z=4$ data are edge-on.
Full line ellipses represent the corresponding $1\times\sigma$ ellipsoids
seen in the same projection. The $3\times\sigma$
ellipse is also plotted for the $z=4$ sample (long dashed line).
The centers of the ellipses are the corresponding projections
of the ellipsoid centers.
Data points for all the massive objects in the samples
are also plotted as circles using darker versions of their respectively colours.
Short dashed lines are the projections of the major axes of the VPs ellipsoids.
 $\stellarmass$ in $M_{\odot}$, $\restar$ in $kpc$ and 
$\sigmastar$ in $km\times s^{-1}$.
\label{fig1}}
\end{figure}

To better view the IDP's, their relation to the VP and any possible evolution we
plot them in Figure~\ref{fig1} for $z=6$ (green), $z=5$ (red), and $z=4$ (blue).
Points are the actual data for all the massive
objects, shown in a projection where the $z=4$ data are edge-on.
Ellipses stand for  the corresponding projections of the 1-sigma
3D ellipsoids  (full lines) or 3-sigma (blue dashed line). The centers of the
ellipses are the corresponding projections of the ellipsoid centers.
Straight lines have the same directions as the major axes of the ellipses
resulting from projections of the VPs ellipsoids.
Two important results arise from this plot and Table~\ref{tab:PCA}.
First, high-$z$ massive galaxies are on IDP's which are  clearly tilted relative to
the VP.
Second, we observe a mild evolution of the IDP between $z=6$ and $z=4$,
primarily driven by changes in the average values $\textit{\~E}_{star}$,
$\textit{\~r}_{star}$, $\textit{\~v}_{star}$ and not the plane parameters
$\alpha^{3D}_{star}$ and $\gamma^{3D}_{star}$.

\section{Discussion \& Conclusions}
To shed more light onto the tilt of the IDP with respect to the VP,
 and following the discussion in $\S$\ref{s1:intro}, we
have first checked if there is a mass homology breaking, that is, if
the parameter $c^{\rm vir}_{\rm M} $ depends on $\stellarmass$.  We have
calculated its trend with stellar mass ($\log \stellarmass \propto
\beta_{vir}^M \log \cmvir$) and listed the best fit
$\beta_{vir}^M$ in Table~\ref{tab:directfits}. Within the error bars
the correlation is consistent with zero.
This means that stars accomodate the product of their spatial
and velocity dispersion distributions (i.e., $\restar \sigmastar$) according to
$\mvir$.
 Second, we have checked whether the mass ratio
$\stellarmass/\mvir$ correlates with $\stellarmass$. And in fact, we
find that this ratio decreases for increasing stellar mass at any
given $z$ (cf. Table~\ref{tab:directfits}). Therefore, we expect the
IDP to be tilted against the VP \citep[as discussed in the
  Introduction and explained in][]{Onorbe:2005,Onorbe:2006}. We
further compared our results against a simulation with different star formation
 parameters and found  no difference concerning the  IDP tilt.

But how can we understand these trends with respect to scenarios of galaxy
formation?
In order to answer this question we need to additionally consider the 
ratio of hot and cold baryon mass inside $\rvir$ (i.e. $\mhbvir/\mcbvir$)
as a function of the mass scale.
First we note that mass assembly of the objects we analyze is dominated by 
cold accretion mode, in consistency with \citet{Keres:2009} results with the entropy-conserving
GADGET-2 code. Now,
 the best-fit parameters to the scaling relation $\log \stellarmass \propto
\beta \log \mhbvir/\mcbvir$ are given in Table~\ref{tab:directfits} again. 
There we find that the fraction of hot over cold baryons increases
very significantly as we go to higher masses. 
We can further acknowledge from Table~\ref{tab:directfits} that the overall
baryon fraction 
$f_B \equiv \mbarvir/\mvir$ does not depend on $\stellarmass$. 
Both these results taken together imply that massive haloes have proportionally
less cold gas available to be accreted from the halo and transformed into stars 
than less massive ones. This explains the
trend of $\stellarmass/\mvir$ with
the mass scale found above.
Further, it is worth noting that all the $\beta$ slopes in
Table~\ref{tab:directfits} change from
one $z$ to another only within their errors.

Now, why is more hot gas relative to cold gas enclosed within the
virial radius as $\mvir$ increases? To answer to this question we have
to recall how massive galaxies assemble their mass. Very briefly, our
simulations show that massive galaxies form from gaseous mass elements
enclosed by overdense subvolumes within the simulation box. 
As predicted by the Adhesion Model \citep{Gurbatov:1989,Vergassola:1994}
we have found that gas is bi-phasic. Indeed,  
at a given time
a distinction can be made between singular gaseous mass elements (as
those that have already been involved in caustic, i.e., singularity,
formation at this time) and regular ones (those that have not yet been
trapped into a caustic and tend to be of low density).
We have also found that,
from a
global point of view, 
mass elements are dynamically organized as a hierarchy of {\it flow convergence
  regions} (FCRs), that is, attraction basins for mass flows. At high
$z$ FCRs undergo fast contractive deformations, 
that violently shrink them,
transforming most of  the cold, densest gaseous mass elements they contain into stars
and heating the diffuse component there. Due to its low density, this component, once heated
tends to keep hot along evolution, and forms shock fronts that expand, in consistency with
\citet{Birnboim:2003} results.  We refer the reader to \citet{DT:2010} for a more elaborate 
discussion.

In the simulations analyzed in this Letter, we have witnessed events
occurring along the fast phase of massive galaxy formation, see
$\S$\ref{s1:intro}: very fast mass assembly, dissipation and star
formation rates ensuing FCR contractive deformations.  These
contractions act on dynamical timescales that are short because we
have high overdensities where massive galaxies are about to form,
therefore explaining the presence of massive objects in a young
Universe.  
Additionally, such violent FCR contractions tend to swallow
the mass close to them, severely limiting the amount of mass available
to be further assembled after they occur.  This would explain why 
a fraction of the objects we have identified are not dynamically disturbed.
 We have also seen in the
simulations the gravitational gas heating due to these violent
dynamical events, that partially transform the ordered mechanical
energy involved in contractions into thermal energy and pressure. This
is a crucial point for understanding the tilt of the IDP's at high
$z$'s. To be quantitative, recall for example that a system must get
rid of an amount of energy equal to its binding energy as it collapses
from infinity and virializes \citep{Binney:2008}. This binding
energy per unit virial mass increases with halo mass as $\mvir^{2/3}$,
so that at assembling a galaxy, the more massive it is the more energy
per unit mass is available to heat and pressurize the gas at the
corresponding FCR contraction.  Otherwise, as explained above, after these
violent events most of the heated low density gas elements remain
hot. This implies that more hot gas relative to cold gas is enclosed
within $\rvir$ as $\mvir$ increases, as we have found.  Therefore, we
can conclude that the origin of the IDPs tilt relative to VPs lies in
that gravitational gas heating processes are more effective as the
mass of the halo increases and that there is not mass homology
breaking.
Finally, let us stress  that the same physical processes act
along the fast phase, as the slopes in  Table~\ref{tab:directfits} do not change  with zs.
%{\bf Again, this interpretation is based upon a
%  detailed analysis of massive galaxy formation 
%  presented in \citet{DT:2010}}

Summing up, the processes involved in high-$z$ massive galaxy
formation are: FCR  contractions (approximatively
equivalent to collapse) 
acting on a bi-phasic gas, induced by singularity formation in terms of
the Adhesion Model approximation; the ensuing transformation of the
ordered mechanical energy of contraction into velocity dispersion, and
then partially into thermal energy and gas pressure, on the same
timescales; and dense gas elements shrinkage, cooling and their
transformation into stars. % Note that the effects of discrete energy
%injection processes have not been explicitly taken into account in the
%simulations we present here, 
%but they have somewhat been mimicked
%through the values of star formation parameters we use. Note, moreover,
%that changing these values does not invalidate
%our conclusion. Anyway, 
Energy injection is
  unlikely to substantially
change the processes responsible for
this high $z$ FP tilt, because, as explained,
 they have to do with caustic
(i.e. singularity) formation.  
We conclude that the violent processes described above
are responsible for having: 1) massive objects at high redhisft, 2)
hot gas coronae, 3) less cold gas to form stars as the mass scale
increases, because we have more gas heated, implying IDP's tilted
relative to VP's, among other results.  We see that the same processes
are responsible for obtaining massive stellar objects shortly after
the Big Bang as well as having them lying on IDPs: fast FCR
contractions at different scales are the engine driving them.

\acknowledgments
We thank the anonymous referee for useful suggestions 
that improved the manuscript.
We thankfully acknowledge to D. Vicente and J. Naranjo for the assistence and technical expertise 
provided at
the Barcelona Supercomputing Centre, as well as the computer resources provided by BSC/RES (Spain).
We thank DEISA Extreme Computing Initiative (DECI) for the CPU time allowed to GALFOBS project.
The Centro de Computaci\'on Cientif\'ica (UAM, Spain) has also provided computing facilities.
This work was partially supported by the DGES (Spain) through the
grants
AYA2009-12792-C03-02/-03, as well as 
by the ASTROMADRID network (CAM S2009/ESP-1496).
AK and JO were supported by the MICINN, Spain, through the Ramon y Cajal programme
and  the ''Supercomputaci\'on y e-Ciencia'' Consolider-Ingenio 2010 project (CSD2007-0050),
respectively.

\begin{deluxetable}{lccccccc}
\tabletypesize{\scriptsize}
\tablecolumns{8}
\tablewidth{0pc}
\tablecaption{Results of PCA analysis.\label{tab:PCA}}
\tablehead{
Sample & No. & $\mbox{\~E}$& $\mbox{\~r}$& $\mbox{\~v}$& $\alpha^{\rm 3D}$& $\gamma^{\rm 3D}$& $\sigma_{\rm Erv}$\\
\cline{1-8}\\
\multicolumn{8}{c}{Halo}
}
\startdata
$z=6$& 137 & 11.321$\pm$0.016 & 1.053$\pm$0.007 & 2.319$\pm$0.006 & 0.930$\pm$0.040 & 1.906$\pm$0.050& 0.0100$\pm$0.0005 \\
$z=5$& 521 & 11.326$\pm$0.010 & 1.118$\pm$0.005 & 2.284$\pm$0.003 & 0.822$\pm$0.017 & 2.008$\pm$0.024& 0.0092$\pm$0.0003 \\
$z=4$& 1315& 11.336$\pm$0.007 & 1.188$\pm$0.003 & 2.248$\pm$0.003 & 0.798$\pm$0.013 & 2.053$\pm$0.018& 0.0105$\pm$0.0002\\
\cutinhead{Stellar}
$z=6$& 137 & 10.189$\pm$0.012 & 0.067$\pm$0.007 & 2.220$\pm$0.006 &-0.142$\pm$0.091 & 2.096$\pm$0.104 & 0.0297$\pm$0.0021\\
$z=5$& 521 & 10.209$\pm$0.007 & 0.119$\pm$0.004 & 2.187$\pm$0.004 &-0.002$\pm$0.059 & 2.040$\pm$0.052 & 0.0290$\pm$0.0010\\
$z=4$& 1315& 10.237$\pm$0.005 & 0.189$\pm$0.003 & 2.146$\pm$0.003 & 0.077$\pm$0.035 & 1.994$\pm$0.027 & 0.0257$\pm$0.0006\\
\enddata
\tablecomments{Column 2: Number of massive galaxies in the sample.
Columns 3, 4 and 5: sample mean values of the $E$, $r$ and $v$ variables (log
$M_{\odot}$, log $kpc$ and log $km\times s^{-1}$ respectively).
Columns 6 and 7: coefficients of the IDP plane.
Column 8: IDP orthogonal scatter in the $E, r$ and $v$ variables.
Errors have beend obtained from a bootstrapping analysis of the samples.}
\end{deluxetable}

\begin{deluxetable}{l|ccc|ccc|}
\tabletypesize{\scriptsize}
\tablecolumns{7}
\tablewidth{0pc}
\tablecaption{Direct Fits.\label{tab:directfits}}
\tablehead{
\multicolumn{1}{c}{$X$} &  \multicolumn{3}{|c|}{$\beta$} & \multicolumn{3}{c|}{$\phi$}\\
\cline{2-7} \\
\colhead{} & \multicolumn{1}{|c}{$z=4$} & \multicolumn{1}{c}{$z=5$} & \multicolumn{1}{c|}{$z=6$} & \multicolumn{1}{c}{$z=4$} & \multicolumn{1}{c}{$z=5$} & \multicolumn{1}{c|}{$z=6$}}
\startdata
$\cf$               &  $-0.071 \pm 0.009$ & $-0.076 \pm 0.014$ & $-0.045 \pm 0.033$ & $1.0173$ & $1.0500$& $0.7267$\\
$\cmvir$    &  $ 0.059 \pm 0.032$ & $ 0.019 \pm 0.066$ & $ 0.074 \pm 0.142$ & $0.8883$ & $1.2765$ & $0.6917$\\
$(\sigmavirial/\sigmastar)^{2}$&  $-0.121 \pm 0.023$ & $-0.131 \pm 0.046$ & $-0.137 \pm 0.096$& $1.4470$ & $1.5277$  & $1.5876$\\
$\reffvirtot/\restar$ & $ 0.252 \pm 0.025$ & $ 0.225 \pm 0.046$ & $ 0.256 \pm 0.110$& $-1.5761$ & $-1.3013$ & $-1.6227$\\
$\stellarmass/\mvir$   & $ -0.306 \pm 0.029$ & $-0.258 \pm 0.057$ & $-0.222 \pm 0.116$& $-2.0301$ & $-1.5159$ & $-1.1264$\\
$\mhbvir/\mcbvir$  & $0.789 \pm 0.022$ & $0.829 \pm 0.040$ & $0.877 \pm 0.091$& $-8.8637$ & $-9.2277$ & $-9.6826$\\
$\mbarvir/\mvir$   & $-0.061 \pm 0.006$ & $-0.048 \pm 0.009$ & $-0.032 \pm 0.142$& $-0.1594$ & $-0.2892$ & $-0.4581$\\
\enddata
\tablecomments{Correlation between various properties $X$ and $\stellarmass$\ as
derived from fitting $\log \stellarmass = \beta \log X + \phi$. Errors stand for
a 97.5\% confidence level intervals.}
\end{deluxetable}

\begin{thebibliography}{}

\bibitem[Bailin et al.(2005)]{Bailin:2005} Bailin, J., et al.\ 
2005, \apjl, 627, L17 

%\bibitem[Binney(1977)]{Binney:1977} Binney, J. 1977, \apj, 215, 483

\bibitem[Binney \& Tremaine(2008)]{Binney:2008} Binney J. \& Tremaine S. 2008, Galactic Dynamics, Princeton 
University Press (Princeton, New Jersey)

%\bibitem[Binney(2004)]{Binney:2004} Binney J. 2004, \mnras, 347, 1093

\bibitem[Birnboim \& Dekel(2003)]{Birnboim:2003} Birnboim, Y., \& Dekel, A.\ 2003, \mnras, 345, 349 

\bibitem[Bouwens et al.(2007)]{Bouwens:2007} Bouwens, R.~J., 
Illingworth, G.~D., Franx, M., \& Ford, H.\ 2007, \apj, 670, 928 

\bibitem[Bouwens et al.(2011)]{Bouwens:2011} Bouwens, R.~J., et al.\ 
2011, \nat, 469, 504 

\bibitem[Bryan \& Norman(1998)]{Bryan:1998} Bryan, G.L. \& Norman, M.L. 1998, ApJ, 495, 80

%%\bibitem[Cimatti et al.(2008)]{Cimatti:2008} Cimatti, A., et al.\ 2008, \aap, 482, 21 

\bibitem[Capak et al.(2011)]{Capak:2011} Capak, P.~L., et al.\ 
2011, arXiv:1101.3586

\bibitem[Cook, Lapi \& Granato(2009)]{Cook:2009} Cook M., Lapi A.,  Granato G. L., 2009, \mnras, 397, 534

%\bibitem[Dickinson(1998)]{Dickinson:1998} {Dickinson} M. 1998, in The Hubble Deep Field, ed. M.~{Livio}, S.~M. {Fall},
%  \& P.~{Madau}, 219--+
  
\bibitem[Dahlen et al.(2010)]{Dahlen:2010} Dahlen, T., et al.\ 
 2010, \apj, 724, 425 



\bibitem[Dom{\'{\i}}nguez-Tenreiro et al.(2006)]{DT:2006} Dom{\'{\i}}nguez-Tenreiro R., O{\~n}orbe J., S{\'a}iz A., Artal H., \& Serna A. 2006, \apjl, 636, L77 

\bibitem[Dom{\'{\i}}nguez-Tenreiro et al.(2010)]{DT:2010} Dom{\'{\i}}nguez-Tenreiro R., O{\~n}orbe J., Mart{\'{\i}}nez-Serrano F.~J. \& Serna A. 2010 \mnras, accepted

\bibitem[Dunkley et al.(2009)]{Dunkley:2009} Dunkley, J., et al. 2009, \apjs, 180, 306

\bibitem[Eggen, Lynden-Bell \& Sandage(1962)]{Eggen:1962} Eggen O.~J., 
Lynden-Bell D., Sandage A.~R., 1962, \apj, 136, 748

%\bibitem[Fardal et al.(2001)]{Fardal:2001} Fardal M.~A., Katz N., Gardner J.~P., Hernquist L., Weinberg D.~H.,
\& Dav{\'e} R.\ 2001, \apj, 562, 605

\bibitem[Giavalisco et~al.(2004)]{Giavalisco:2004} {Giavalisco} M., {Dickinson} M., {Ferguson} H.~C., {et~al.}
  2004, \apjl, 600, L103

\bibitem[Gurbatov et al.(1989)]{Gurbatov:1989} Gurbatov S.~N.,
Saichev A.~I.,  Shandarin S.~F., 1989, \mnras, 236, 385

%\bibitem[Katz et al.(2003)]{Katz:2003} Katz N., Keres D., Dave R.,
%\& Weinberg D.~H.\ 2003, The IGM/Galaxy Connection.~The Distribution of Baryons at z=0, 281, 185

%\bibitem[Kay et al.(2000)]{Kay:2000} Kay, S.~T., Pearce, F.~R., Jenkins, A., Frenk, C.~S., White, S.~D.~M., Thomas, P.~A., \& Couchman, H.~M.~P.\ 2000, \mnras, 316, 374

\bibitem[Kere{\v s} et al.(2009)]{Keres:2009} Kere{\v s}, D., Katz, N., Fardal, M., Dav{\'e}, R., \& Weinberg, D.~H.\ 2009, \mnras, 395, 160

\bibitem[Knollmann 
\& Knebe(2009)]{Knollmann:2009} Knollmann, S.~R., \& Knebe, A.\ 2009, \apjs, 182, 608 

%%\bibitem[Kriek et al.(2008)]{Kriek:2008} Kriek, M., van der Wel, A., van Dokkum, P.~G., Franx, M., 
%%\& Illingworth, G.~D.\ 2008, \apj, 682, 896 

%%\bibitem[Komatsu et al.(2010)]{Komatsu:2010} Komatsu, E., et al.\ 
%%2010, arXiv:1001.4538 

\bibitem[Larson(1974)]{Larson:1974} Larson R.~B., 1974, \mnras, 
166, 585 

\bibitem[Mancini et 
al.(2009)]{Mancini:2009} Mancini, C., Matute, I., Cimatti, A., Daddi, E., Dickinson, M., Rodighiero, G., Bolzonella, M., \& Pozzetti, L.\ 2009, \aap, 500, 705 

\bibitem[Mart{\'{\i}}nez-Serrano et al.(2008)]{MartinezSerrano:2008} 
Mart{\'{\i}}nez-Serrano, F.~J., Serna, A., Dom{\'{\i}}nguez-Tenreiro, R., 
\& Moll{\'a}, M.\ 2008, \mnras, 388, 3

\bibitem[McLure et~al.(2006)]{McLure:2006}
{McLure}, R.~J., {Cirasuolo}, M., {Dunlop}, J.~S., {et~al.} 2006, \mnras, 372,
  357

\bibitem[Mobasher et al.(2005)]{Mobasher:2005} Mobasher B. et al., 2005, ApJ, 635, 832

\bibitem[Mobasher \& Wiklind(2010)]{Mobasher:2010} Mobasher B., Wiklind T., 2010, in  The Impact of HST on European Astronomy, Astrophysics and Space Science Proceedings, F. Duccio Macchetto ed., Springer, Netherlands

\bibitem[O\~norbe et al.(2005)]{Onorbe:2005} O\~norbe, J., Dom\'{\i}nguez-Tenreir,o R., S\'aiz, A., Serna, A., Artal, H.\ 2005, \apj, 632, L57

\bibitem[O\~norbe et al.(2006)]{Onorbe:2006} O\~norbe, J., Dom\'{\i}nguez-Tenreiro, R., S\'aiz, A., Artal, H., Serna, A.\ 2006, \mnras, 373, 503

\bibitem[Oser et al.(2010)]{Oser:2010} Oser, L., Ostriker, J.~P., 
Naab, T., Johansson, P.~H., \& Burkert, A.\ 2010, arXiv:1010.1381 

\bibitem[Padmanabhan(1993)]{Padmanabhan:1993} Padmanabhan, T.\ 1993, 
Structure Formation in the Universe, by T.~Padmanabhan, pp.~499.~ISBN 
0521424860.~Cambridge, UK: Cambridge University Press, June 1993.

%\bibitem[Rees \& Ostriker(1977)]{Rees:1977} Rees, M.~J., \& Ostriker, J.~P.\ 1977, \mnras, 179, 541 

\bibitem[Rodighiero et 
al.(2007)]{Rodighiero:2007} Rodighiero, G., Cimatti, A., Franceschini, A., Brusa, M., Fritz, J., \& Bolzonella, M.\ 2007, \aap, 470, 21 

%%\bibitem[Saglia et al.(2010)]{Saglia:2010} Saglia, R.~P., et al.\ 2010, \aap, 524, A6 

\bibitem[S\'aiz et al. (2004)]{Saiz:2004} S\'aiz, A., Dom{\'{\i}}nguez-Tenreiro, R., \& Serna, A. 2004, \apj, 601L, 131

\bibitem[Salvador-Sol\'e et al.(2005)]{SalvadorSole:2005} Salvador-Sol\'e E.,  Manrique A., Solanes J.~M. 2005,
\mnras, 358, 901

\bibitem[Serna, Dom{\'{\i}}nguez-Tenreiro, {\&} S\'aiz(2003)]{Serna:2003} Serna, A., Dom{\'{\i}}nguez-Tenreiro, R., \& S\'aiz, A. 2003, \apj, 597, 878

%\bibitem[Slyz et al.(2005)]{Slyz:2005} Slyz A.~D., Devriendt J.~E.~G., Bryan G., Silk J.,\ 2005, \mnras, 356, 737

\bibitem[Stark et al. (2009)]{Stark:2009} Stark, D.P., Ellis, R.S., Bunker, A., Bundy, K., Targett, T., Benson, A., Lacy.,\ 2009, \apj, 697, 149


%\bibitem[Steidel et~al.(1999)]{Steidel:1999}
%{Steidel}, C.~C., {Adelberger}, K.~L., {Giavalisco}, M., {Dickinson}, M., \&
%  {Pettini}, M. 1999, \apj, 519, 1
  
%\bibitem[Steidel et~al.(2003)]{Steidel:2003}
%{Steidel}, C.~C., {Adelberger}, K.~L., {Shapley}, A.~E., {et~al.} 2003, \apj,
%  592, 728

%\bibitem[Tinsley(1972)]{Tinsley:1972} Tinsley B.~M.,\ 1972, \apj, 
%178, 319 

\bibitem[Toomre(1977)]{Toomre:1977} Toomre A., 1977, in The Evolution of Galaxies and Stellar Populations, eds.\ B. Tinsley \& R. Larson (New Have, CN: Yale Univ.\ Press)

%%\bibitem[Trujillo et al.(2007)]{Trujillo:2007} Trujillo, I., Conselice, C.~J., Bundy, K., Cooper, M.~C., Eisenhardt, P., \& Ellis, R.~S.\ 2007, \mnras, 382, 109 

%%\bibitem[van Dokkum et al.(2008)]{vanDokkum:2008} van Dokkum, P.~G., et al.\ 2008, \apjl, 677, L5 

\bibitem[Vergassola et al.(1994)]{Vergassola:1994} Vergassola~M., Dubrulle~B., Frisch~U., Noullez~A., 1994, A\&A, 289, 325

\bibitem[Wechsler et al.(2002)]{Wechsler:2002}
Wechsler R.H., Bullock J.S., Primack J.R., Kravtsov A.V., Dekel A., 2002, ApJ, 568, 52

\bibitem[Weinberg et al.(1997)]{Weinberg:1997} Weinberg, D.~H., 
Hernquist, L., \& Katz, N.\ 1997, \apj, 477, 8 

\bibitem[White \& Rees(1978)]{White:1978} White, S.~D.~M., \& Rees, M.~J.\ 1978, \mnras, 183, 341 

\bibitem[Wiklind et~al.(2008)]{Wiklind:2008}
{Wiklind}, T., {Dickinson}, M., {Ferguson}, H.~C., {et~al.} 2008, \apj, 676,
  781

\bibitem[Yan et~al.(2006)]{Yan:2006}
{Yan}, H., {Dickinson}, M., {Giavalisco}, M., {et~al.} 2006, \apj, 651, 24

\bibitem[Zhao et al.(2003)]{Zhao:2003}
Zhao D.H., Mo H.J., Jing Y.P.,  Borner G., 2003, \mnras, 339, 12


\end{thebibliography}
\end{document}